\documentclass[11pt]{article}
\usepackage{graphicx} 
\usepackage{amsmath}
\usepackage{amsthm}
\usepackage{amssymb}
\usepackage{colonequals}
\usepackage[margin=1in]{geometry}
\usepackage{enumitem}
\usepackage{subcaption}
\usepackage{float}
\usepackage{xcolor}

\usepackage{siunitx}
\usepackage{rotating}
\usepackage{makecell}
\usepackage{bm}

\newtheorem{theorem}{Theorem}[section]

\newtheorem{definition}[theorem]{Definition}
\newtheorem{example}[theorem]{Example}

\newtheorem{conjecture}[theorem]{Conjecture}
\theoremstyle{plain} 
\newcommand{\thistheoremname}{}
\newtheorem*{genericthm}{\thistheoremname}

\theoremstyle{definition}

\theoremstyle{plain}

\usepackage{hyperref}
\definecolor{cblack}{rgb}{0,0,0}
\definecolor{cblue}{rgb}{0.121569,0.466667,0.705882}    
\definecolor{corange}{rgb}{1.000000,0.498039,0.054902}  
\definecolor{cgreen}{rgb}{0.172549,0.627451,0.172549}   
\definecolor{cred}{rgb}{0.839216,0.152941,0.156863}     
\definecolor{cpurple}{rgb}{0.580392,0.403922,0.741176}  
\definecolor{cbrown}{rgb}{0.549020,0.337255,0.294118}   
\definecolor{cpink}{rgb}{0.890196,0.466667,0.760784}
\definecolor{cgray}{rgb}{0.498039,0.498039,0.498039}
\definecolor{cgreen2}{rgb}{0.7372549019607844, 0.7411764705882353, 0.13333333333333333}

\definecolor{clightgray}{rgb}{0.6,0.6,0.6}
\definecolor{cllightgray}{rgb}{0.8,0.8,0.8}
\hypersetup{
  linkcolor  = cblue,
  citecolor  = cgreen,
  urlcolor   = corange,
  colorlinks = true,
}

\DeclareMathOperator*{\argmax}{arg\,max}
\DeclareMathOperator*{\argmin}{arg\,min}

\newcommand{\greedy}{\mathrm{grd}}
\newcommand{\reluctant}{\mathrm{rel}}
\newcommand{\Exp}{\mathrm{Exp}}
\newcommand{\Unif}{\mathrm{Unif}}

\newcommand{\supp}{\mathrm{supp}}
\newcommand{\Sparsify}{\mathrm{Sparsify}}
\newcommand{\Rad}{\mathrm{Rad}}

\newcommand{\PP}{\mathbb{P}}
\newcommand{\EE}{\mathbb{E}}

\newcommand{\sN}{\mathcal{N}}

\newcommand{\what}{\widehat}
\renewcommand{\hat}{\widehat}

\title{Empirical universality and non-universality of local dynamics in the Sherrington-Kirkpatrick model}
\date{March 8, 2026}
\usepackage{authblk}
\author{Grace Liu\thanks{Email: \texttt{gliu44@jhu.edu}}\,\,}
\author{Dmitriy Kunisky\thanks{Email: \texttt{kunisky@jhu.edu}.}}
\affil{Department of Applied Mathematics \& Statistics, Johns Hopkins University}

\begin{document}

\maketitle
\thispagestyle{empty}

\begin{abstract}
    Several recent works have aimed to design algorithms for optimizing the Hamiltonians of spin glass models from statistical physics.
    While Montanari~(2018) eventually gave a sophisticated message-passing algorithm to do this nearly optimally for the Sherrington-Kirkpatrick~(SK) model, the recent work of Erba, Behrens, Krzakala, and Zdeborov\'{a}~(2024) also observed that a simple yet unusual algorithm first proposed by Parisi~(2003) seems to perform just as well: perform local \emph{reluctant} search, repeatedly making the local adjustment improving the objective function by the smallest possible amount.
    This is in contrast to the more intuitive local \emph{greedy} search that repeatedly makes the local adjustment improving the objective by the largest possible amount.
    We study empirically how the performance of these algorithms depends on the distribution of entries of the coupling matrix in the SK model.
    We find evidence that, while the runtime of greedy search enjoys universality over a broad range of distributions, the runtime of reluctant search surprisingly is \emph{not} universal, sometimes depending quite sensitively on the entry distribution.
    We propose that one mechanism leading to this non-universality is a change in the behavior of reluctant search when the couplings have discrete support on an evenly-spaced grid, and give experimental results supporting this proposal and investigating other properties of a distribution that might affect the performance of reluctant search.
\end{abstract}

\clearpage

\tableofcontents
\thispagestyle{empty}

\clearpage

\section{Introduction}
\pagenumbering{arabic}

\subsection{Ising and Sherrington-Kirkpatrick Models}

In this paper, we will study instances of the \emph{Ising model}, a classical model of magnetism in statistical physics; see, e.g., the books \cite{MPV-1987-SpinGlassTheoryBeyond,Nishimori-2001-SpinGlassesInformationProcessing,MM-2009-InformationPhysicsComputation,charbonneau2023spin} for discussion of the physics perspective on the models we will consider.
A general version of the Ising model is specified by a Hamiltonian
\[H(J, \sigma) = -\frac{1}{2}\sigma^{\top} J \sigma = -\frac{1}{2}\sum_{i, j = 1}^N J_{ij} \sigma_i \sigma_j\]
for a symmetric matrix $J \in \mathbb{R}^{N \times N}$ called the \emph{coupling matrix}.
We view this as a function of $\sigma = (\sigma_1, \dots, \sigma_N) \in \{\pm 1\}^N$, parametrized by $J$.
Since $\sigma_i^2 = 1$ does not depend on $\sigma$, we assume that $J_{ii} = 0$; if not, then all values of $H$ are merely shifted by the constant $\frac{1}{2}\sum_{i = 1}^N J_{ii}$, which would be inconsequential for our purposes.

The original Ising model took $J$ to be the adjacency matrix of a graph such as a finite lattice, or its negative, to model ferromagnetic or antiferromagnetic interactions of the \emph{spins} $\sigma_i$, respectively.
Taking $J$ to have entries with \emph{random} (or in general ``incoherent'' in an appropriate sense) signs leads to models of \emph{spin glasses}, disordered magnetic materials often having more complex phenomenology than ordinary magnets.
A rich theory has emerged around the behavior of these more complex models, describing how \emph{frustration}---the impossibility of making the signs of all or most terms of the Hamiltonian agree---leads to rough energy landscapes and slower relaxation to equilibrium.
See, e.g., \cite{Anderson-1988-SpinGlass,stein2013spin} for expository accounts.
Because of these features, spin glass Hamiltonians have also been proposed as a suitable model for difficult optimization problems, even ones stemming from fields other than physics.
Exploring this analogy has led, for example, to useful insights about the structure of hard constraint satisfaction problems and the statics and dynamics of training large machine learning models \cite{krzakala2007gibbs,ding2015proof,ZK-2016-Review,Montanari-2024-RandomOptimizationStatisticalPhysics}.

Perhaps the simplest spin glass model takes the entries of $J$ to be independent and identically distributed.
We give such random matrices the following name.
\begin{definition}
    \label{def:wigner}
    We call $J \in \mathbb{R}^{N \times N}$ a \emph{normalized Wigner matrix} with entry distribution $\mu$ if $J_{ii} = 0$ for all $i \in [N]$ and $\sqrt{N} \cdot J_{ij} \sim \mu$ are i.i.d.\ and $J_{ij} = J_{ji}$ for all $i, j \in [N]$ with $i < j$.
\end{definition}
\noindent
The reason for the choice of scaling $J_{ij} = O(N^{-1/2})$ is that, under mild regularity conditions on $\mu$, the operator norm $\|J\|$ then has order $O(1)$ with high probability as $n \to \infty$ (by the results of \cite{bai1988necessary}, it suffices just for $\mu$ to have a finite fourth moment for this to hold; see the book \cite{AGZ-2010-RandomMatrices} for general discussion).

Both this entire class or the special case of i.i.d.\ Gaussian entries are called the \emph{Sherrington-Kirkpatrick~(SK) model(s)} after the pioneering work of \cite{SK-1975-SolvableModel}.
Even among spin glasses, the SK model and the energy landscape induced by its Hamiltonian have an exceptionally rich structure, consisting of a nested geometric structure of energy level sets described by \emph{ultrametricity} and \emph{(full) replica symmetry breaking} in the seminal theory proposed by Parisi based on non-rigorous methods of statistical physics and eventually validated in the mathematics literature \cite{Parisi-1979-SK,Parisi-1980-SequenceApproxSK,Guerra-2003-BrokenRSB,Talagrand-2006-Parisi,Talagrand-2010-MeanFieldSpinGlasses1,Panchenko-2013-SK}.
The SK model is a \emph{mean field} spin glass model, meaning that it involves every particle $\sigma_i$ interacting with every other particle (by appearing together with every other particle in a term of the Hamiltonian).
Other cases of the SK model with different entrywise distributions are similarly dense, unless the distribution of $J_{ij}$ takes the value zero with very high probability.

Here we study certain properties of the random optimization problem associated to the Hamiltonian of the Sherrington-Kirkpatrick model, following another line of work in the probability and statistical physics literature \cite{ABM-2018-CREMOptimization,Subag-2018-FullRSBOptimization,Montanari-2019-SKOptimization,AMS-2020-MeanFieldOptimization}.
In this problem, we wish to find the vector $\sigma 
\in \{\pm 1\}^N$ that minimizes $H(J, \sigma)$ for a given $J$.
In physical language, the optimizer is called the \emph{ground state configuration} of the model, while its objective value is called the \emph{ground state energy}.

We explore the natural probabilistic property of \emph{universality} of the behavior of algorithms for performing this task.
As in many settings in probability theory, one perhaps expects the behavior of simple algorithms to not depend very much on the specific distribution of $J$, as, for instance, the large-scale behavior of sums of i.i.d.\ random variables does not depend on their distribution in the central limit theorem, or the large-scale behavior of a random walk does not depend on the step distributions in Donsker's invariance principle in classical probability (see any standard reference such as \cite{Durrett-1995-ProbabilityTheoryExamples,Klenke-2014-ProbabilityTheory}).
But, the main purpose of this paper is to give empirical evidence and partial theoretical arguments that, surprisingly, for a particular algorithm of recent interest in the literature, universality \emph{fails}, and its behavior depends, sometimes quite sensitively, on the entrywise distribution of $J$.

\subsection{Local and Non-Local Algorithms}

We focus on \emph{local algorithms} for heuristically solving the optimization problem associated to the SK model (that is, without guarantees of necessarily reaching the global optimizer, which is an NP-hard problem for general $J$ as in the special case of $J$ a graph Laplacian it can encode the \emph{maximum cut} problem \cite{Karp-1972-Reducibility}).
A local algorithm is an iterative one that makes changes to $\sigma$ that are not too large in the following metric.
\begin{definition}[Hamming distance]
    The \emph{Hamming distance} on $\{\pm 1\}^N$ is
    \[ d(\sigma, \sigma^{\prime}) \colonequals \#\{i: \sigma_i \neq \sigma_i^{\prime}\}. \]
\end{definition}
\noindent
A local algorithm repeatedly changes $\sigma$ to some $\sigma^{\prime}$ such that $d(\sigma, \sigma^{\prime}) \leq K$ for some reasonably small $K$.\footnote{Note that, for any given $\sigma \in \{\pm 1\}^N$, most other configurations are at Hamming distance $\Theta(N)$, so local algorithms are severely constrained in the kinds of updates they can make provided that $K \ll N$.}
We will focus specifically on the case $K = 1$, meaning that our algorithms can only change one coordinate of $\sigma$ at a time.

More formally, we may view a local algorithm as producing a sequence $\sigma^{(t)}$ starting from a given initialization $\sigma^{(0)}$.
In general, there can be an arbitrary rule for constructing the next iterate,
\[ \sigma^{(t + 1)} = F(\sigma^{(t)}, J), \]
provided that $d(\sigma, F(\sigma, J)) \leq 1$ for all $\sigma$.
When $\sigma^{(T + k)} = \sigma^{(T)}$ for all $k \geq 1$, we say that the local algorithm has \emph{converged} in $T$ iterations.
As a final twist, we allow \emph{randomized local algorithms}, where instead
\[ \sigma^{(t + 1)} = F(\sigma^{(t)}, J, z^{(t)}) \]
depends on some extra random variable $z^{(t)}$, for a collection of such random variables that are i.i.d.\ and independent of $J$.
For instance, one may use this extra randomness to implement local Markov Chain Monte Carlo approaches like Metropolis dynamics as a local algorithm in our sense (see \cite{MY-1983-StaticsDynamicsIsingSK,billoire2011dynamics} for the definition and discussion of the behavior of these dynamics in our setting).

While we do not consider it here, also worth mentioning is the slightly different framework of \emph{sequential local algorithms} studied by \cite{dandi2025sequential} (one instance of which was also compared to the algorithms we do focus on by \cite{Parisi-2003-ZeroTemperatureSK}).
These algorithms repeatedly scan through the spins in a fixed order and update them according to local information around the spin under consideration (in particular, in that work, according to rules involving the local \emph{field} acting on spin $i$, which is the quantity $(J\sigma)_i$).
Though this is a quite restrictive structure to demand of an algorithm, \cite{dandi2025sequential} show that such algorithms, unlike (at least with current tools) those we propose to study empirically below, are amenable to rigorous theoretical analysis.

We will focus instead on a family of even simpler local algorithms, which we describe next.
Though they may seem naive, we will see that even among these algorithms there are interesting and surprising behaviors.

\paragraph{Greedy algorithm}
The most intuitive local algorithm is the \emph{greedy algorithm}.
The greedy algorithm makes the largest possible local improvement to the objective function, flipping the spin making the largest reduction in the system's energy.
Formally, its update rule is
\[ F_{\greedy}(\sigma, J) \colonequals \argmin_{\substack{\sigma^{\prime} \in \{\pm 1\}^N \\ d(\sigma, \sigma^{\prime}) \leq 1}} H(\sigma^{\prime}, J). \]
One may view this as a version of gradient descent in our discrete setting.
As in many settings in optimization, such an algorithm does not necessarily perform well on optimization landscapes with many local minima, where it can get stuck.
Specifically in the case of the SK model, this issue becomes increasingly pronounced as the system size $N$ grows, since the number of local minima at energy levels far from the ground state energy grows as $\exp(\Omega(N))$, as shown by \cite{TE-1980-SKGroundState,ABDLO-2019-SKLocalOptima}.

\paragraph{Efficient non-local improvements}
It turns out that, despite the complexity of its optimization landscape, there are efficient algorithms that can nearly perfectly optimize the Hamiltonian of the SK model, with high probability outputting a point with energy that is, say, 99.9\% of the ground state energy.
Unfortunately, the only algorithm that has been rigorously proven to achieve such a guarantee is a much more complicated one (in particular, not a local algorithm) motivated by approximate message passing \cite{Montanari-2019-SKOptimization}.
It is natural to ask: is this complexity really needed, or can local algorithms be ``redeemed'' and still perform well on the SK model's Hamiltonian?
While, to the best of our knowledge, it is not rigorously known that the greedy algorithm does not achieve this, the above results as well as other heuristic theoretical analysis and numerical experiments strongly suggest that this is the case (see the results and statistical physics analysis of \cite{Parisi-2003-ZeroTemperatureSK,BCDG-2003-SKEnergyDecreasingDynamics,Horner-2007-SKLocal,EBKZ-2024-QuenchesSKReluctant}).

\paragraph{Reluctant algorithm}
Still, evidence has emerged that the answer to the above question might be \emph{yes}, and using an algorithm that is just as simple as the greedy algorithm.
However, what seems to be the correct choice of algorithm appears quite bizarre at first: this \emph{reluctant algorithm} does just the \emph{opposite} of the greedy algorithm, flipping the spin that improves the objective function by the \emph{least} amount.
Formally, its update rule is
\[ F_{\reluctant}(\sigma, J) \colonequals \argmax_{\substack{\sigma^{\prime} \in \{\pm 1\}^N \\ d(\sigma, \sigma^{\prime}) \leq 1 \\ H(\sigma^{\prime}, J) < H(\sigma, J)}} H(\sigma^{\prime}, J). \]
We adopt the convention that, if there are no feasible $\sigma^{\prime}$ for the $\argmax$, then $F_{\reluctant}(\sigma, J)$ returns $\sigma$.
A reasonable intuition for why this is a good idea is that it is advantageous to ``slow down'' greedy descent on a complicated optimization landscape, in order for the trajectory of iterates to have the opportunity to explore more of the landscape.
Empirical studies indeed have verified that the reluctant algorithm consistently finds lower-energy configurations than the greedy algorithm \cite{Parisi-2003-ZeroTemperatureSK,BCDG-2003-SKEnergyDecreasingDynamics,Horner-2007-SKLocal}, and, while at first smaller experiments (in particular those carried out by \cite{Parisi-2003-ZeroTemperatureSK}) suggested otherwise, the recent work of \cite{EBKZ-2024-QuenchesSKReluctant} found that both larger experiments and non-rigorous analysis using the methods of statistical physics in fact indicate that the reluctant algorithm might with high probability find configurations with energy arbitrarily close to the ground state energy.
This would mean that the very simple reluctant local algorithm indeed matches the performance of the far more complicated algorithm of \cite{Montanari-2019-SKOptimization}.

\paragraph{$\lambda$-Reluctant algorithm}
Finally, \cite{contucci2005interpolating} propose a family of algorithms that interpolates between greedy and reluctant dynamics, allowing one to choose (intuitively speaking) from a continuous range of options trading off between fast convergence and thorough exploration of the energy landscape.
These algorithms depend on a parameter $\lambda \in (0, +\infty)$ and are randomized local algorithms, their update rule taking as input the ancillary random variables $z^{(t)} \sim \Exp(1)$.
Actually, the algorithm itself works with $z^{(t)} / \lambda$, whose law is $\Exp(\lambda)$.
The algorithm then proceeds by choosing the spin to flip that reduces the energy by an amount closest to $z^{(t)} / \lambda$.
Formally, the random update rule is
\[ F_{\lambda}(\sigma, J, z^{(t)}) \colonequals \argmin_{\substack{\sigma^{\prime} \in \{\pm 1\}^N \\ d(\sigma, \sigma^{\prime}) \leq 1 \\ H(\sigma^{\prime}, J) < H(\sigma, J)}} \left|H(\sigma, J) - H(\sigma^{\prime}, J) - \frac{z^{(t)}}{\lambda}\right|. \]
As $\lambda \to 0$, $z^{(t)} / \lambda \to \infty$, so the minimum above is achieved by the $\sigma^{\prime}$ giving the largest reduction in energy, recovering the greedy algorithm.
As $\lambda \to \infty$, $z^{(t)} / \lambda \to 0$, so the minimum is achieved by the $\sigma^{\prime}$ giving the smallest reduction in energy (but still constrained to reduce the energy), recovering the reluctant algorithm.
For this reason we will refer to the greedy algorithm as the case $\lambda = 0$ and the reluctant algorithm as the case $\lambda = +\infty$ of the $\lambda$-reluctant algorithm.

\subsection{Runtime Scaling Laws}

It is in the behavior of the reluctant and $\lambda$-reluctant algorithms that we will give evidence of non-universality.
This non-universality pertains to the scaling obeyed by the runtime of this algorithm, where by \emph{runtime} we mean the smallest step index $T$ at which the algorithm has converged.
Let us describe what prior work suggests about this scaling.

The empirical work of \cite{contucci2005interpolating} gave strong evidence for the following conjecture, that the runtime of any $\lambda$-reluctant algorithm, including the extremal greedy and reluctant algorithms, scales polynomially in $N$ to leading order for the classical SK model with $J$ having i.i.d.\ Gaussian entries.
We reproduce some of their experiments here in Section~\ref{sec:results1-scaling}, and also repeat these experiments with different entry distributions $\mu$.
We find evidence for the following slightly stronger proposal, that in fact the runtime is polynomial for \emph{any} reasonable entry distribution.

\begin{conjecture}[Informal]
    \label{conj:power-law}
    For any ``sufficiently nice'' probability measure $\mu$ on $\mathbb{R}$ with $\mathbb{E}_{X \sim \mu} X = 0$ and $\mathbb{E}_{X \sim \mu} X^2 = 1$ and any $\lambda \in [0, +\infty]$, if $\sqrt{N} \cdot J$ has i.i.d.\ entries distributed as $\mu$, if $T = T(N, \mu, \lambda)$ is the (random) runtime of reluctant dynamics on an $N \times N$ random matrix $J$, then there exist constants $\alpha = \alpha(\mu, \lambda) > 0$ and $\beta = \beta(\mu, \lambda) > 0$ such that the convergence
    \begin{equation}
    \frac{T(N, \mu, \lambda)}{\alpha(\mu, \lambda)N^{\beta(\mu, \lambda)}} \to 1.
    \label{eq:power-law}
    \end{equation}
    holds in probability as $N \to \infty$.
\end{conjecture}

Less precisely, the claim says that the leading order behavior of the runtime is
\[ T(N, \mu, \lambda) \approx \alpha(\mu, \lambda) N^{\beta(\mu, \lambda)}. \]
The only truly informal part of the conjecture is that we are unsure of regularity conditions on $\mu$ that should be necessary for the claim to hold.
Our experiments will show that, at least, the conjecture holds for a wide range of subgaussian $\mu$, so perhaps subgaussianity is a reasonable conservative notion of regularity to substitute into the statement.

We note, anticipating the setup of our numerical experiments, that an equivalent way to state the convergence \eqref{eq:power-law} is
\begin{equation}
    \label{eq:log-power-law}
    \big|\log T(N, \mu, \lambda) -\beta(\mu, \lambda) \log N - \log \alpha(\mu, \lambda)\big| \to 0.
\end{equation}
Thus, we expect that the familiar technique of performing linear regression on a log-log plot should apply to estimate the constants $\alpha$ and $\beta$.\footnote{We note that the influential work \cite{CSN-2009-PowerLawsEmpirical} pointed out important flaws in this common methodology, but, as we will see, the agreement of our synthetic data with the fitted power laws is so strong that we do not expect those caveats to apply here.}

\subsection{Summary of Results: Universality and Non-Universality}

Universality is a phenomenon often observed in the study of large random systems, where key large-scale behaviors of the system depend only on a few statistical features of the underlying randomness, and not on the details of its entire distribution.
A classical example is the central limit theorem: the asymptotic Gaussian behavior of a normalized sum of i.i.d.\ random variables only depends on their having mean 0 and variance 1, and otherwise is completely robust to the choice of underlying distribution.
A more modern example is Wigner's semicircle limit theorem: for any $J = J^{(N)}$ as in Definition~\ref{def:wigner}, the distribution of eigenvalues of $J$ converges as $N \to \infty$ (in a suitable sense) to the universal \emph{semicircle distribution} provided only that the entry distribution $\mu$ has mean 0 and variance 1 (many proofs are given, for example, in \cite{AGZ-2010-RandomMatrices}).

The same occurs in more complicated disordered systems like the SK model.
For instance, the SK model's \emph{free energy} (a certain powerful functional summary statistic of the model's behavior and energy landscape) is universal across a wide class of entry distributions including Gaussian, Rademacher, and others \cite{carmona2004universalitysherringtonkirkpatricksspinglass}. 
On the other hand, universality is not always guaranteed to hold.
As an example from statistical physics, the work \cite{lundow2014isingspinglassdimension} demonstrated that the critical behavior of the Ising spin glass on a four-dimensional lattice (a much sparser version of the SK model with non-zero couplings associated to the edges of this graph) \emph{does} depend on the distribution of couplings \(J\), with Gaussian, Laplacian, and bimodal entry distributions giving different scalings of certain associated quantities.

We are now ready to pose precisely the question we will study in this paper:
\begin{gather*} 
\text{Is the scaling exponent $\beta(\mu, \lambda)$ of $\lambda$-reluctant dynamics'} \\
\text{runtime universal over the entry distribution $\mu$?}
\end{gather*}
We will always assume in such discussion, as in the above examples of universality, that $\mu$ is normalized to have mean 0 and variance 1.
We also note that we ask this question for each value of $\lambda$ (which specifies one of the $\lambda$-reluctant algorithms).

On the one hand, we will find indications that the case $\lambda = 0$ of greedy algorithms does appear to behave universally over a wide range of entry distributions $\mu$.
Those results suggest the following conjecture.
\begin{conjecture}[Informal]
    \label{conj:greedy-univ}
    $\beta(\mu, 0) \approx \num{1.1}$ is a constant independent of $\mu$ for any ``sufficiently nice'' $\mu$ with mean 0 and variance 1.
\end{conjecture}

On the other hand, we will see that once $\lambda > 0$, then the $\lambda$-reluctant algorithms appear not to behave universally.
While we will present some results for finite positive $\lambda$, let us focus here on the case $\lambda = \infty$ of reluctant dynamics, for which our results are clearer and easier to present.
We will identify the following property of the \textbf{discrepancy} of $\mu$ as having a considerable effect on $\beta(\mu, \infty)$.

\begin{definition}[Discrepancy of a probability measure]
    \label{def:discrepancy}
    For a probability measure $\mu$ supported on $\supp(\mu) \subseteq \mathbb{R}$, we define its \emph{discrepancy} to be the quantity
    \[ \Delta(\mu) \colonequals \inf_{n \geq 1} \inf_{\substack{x_1, \dots, x_n \in \supp(\mu) \\ s_1, \dots, s_n \in \{\pm 1\} \\ \sum_{i = 1}^n s_i x_i > 0}} \sum_{i = 1}^n s_i x_i = \inf_{n \geq 1} \inf_{\substack{x_1, \dots, x_n \in \supp(\mu) \\ s_1, \dots, s_n \in \{\pm 1\}}} \left|\sum_{i = 1}^n s_i x_i\right|.  \]
\end{definition}
\noindent
While the notation is heavy, the idea is simply that $\Delta(\mu) > 0$ if $\mu$ is supported on a ``grid'' of real numbers.
The following examples illustrate how the structure of the support determines whether the discrepancy is positive or zero.

\begin{example}[Continuous support]
If \(0\) is a limit point of \(\supp(\mu)\), then for every \(\epsilon > 0\), we can find some \(x \in \supp(\mu)\) with \(|x| < \epsilon\). 
Taking \(n = 1\) and \(s_1 = \pm 1\), the term \(s_1 x_1\) itself can be made arbitrarily close to zero, thus the infimum over all such sums is zero.
For example, the standard Gaussian distribution has \(\Delta(N(0, 1)) = 0\).
\end{example}

\begin{example}[Integer support]
If \(\supp(\mu) \subset \mathbb{Z}\) is finite and \(\gcd(\supp(\mu)) = d\), then every signed sum \(\sum_i s_i x_i\) is an integer multiple of \(d\). 
In fact, by the solution of a linear Diophantine equation in elementary number theory, there exists such a signed sum equal to $d$, so $\Delta(\mu) = d > 0$.
For instance, $\Delta(\Unif(\{-1, 1\}) = \Delta(\{2, 3\}) = 1$, while $\Delta(\Unif(\{2, 4\}) = 2$.
\end{example}

\begin{example}[Discrete support with discrepancy zero]
On the other hand, discrepancy zero is not the same as having discrete support.
If the support of \(\mu\) contains two nonzero numbers whose ratio is irrational, then, since integer combinations of these numbers are dense in \(\mathbb{R}\), $\Delta(\mu) = 0$.
For instance, this applies to $\mu = \Unif(\{1, \sqrt{2}\})$ and similar constructions.
\end{example}

Our results suggest that the distributions with positive discrepancy form a universality class, in the sense that they share a scaling exponent.
\begin{conjecture}[Informal]
    \label{conj:delta-pos-univ}
    $\beta(\mu, \infty) \approx \num{1.6}$ is a constant independent of $\mu$ for any ``sufficiently nice'' $\mu$ with mean 0, variance 1, and discrepancy $\Delta(\mu) > 0$.
\end{conjecture}
\noindent
However, we believe that distributions with zero discrepancy (such as ones absolutely continuous with respect to Lebesgue measure) both do not fall into this universality class and do not themselves form a single second universality class.
\begin{conjecture}
    \label{conj:delta-zero-nonuniv}
    There exist $\mu_1, \mu_2, \mu_3$ all having mean 0 and variance 1 and such that $\Delta(\mu_1) > 0$ while $\Delta(\mu_2) = \Delta(\mu_3) = 0$, such that the numbers $\beta(\mu_1, \infty)$, $\beta(\mu_2, \infty)$, and $\beta(\mu_3, \infty)$ are all distinct.
\end{conjecture}
\noindent
More speculatively, it is consistent with our results that all $\mu$ with $\Delta(\mu) = 0$ have \emph{greater} exponent $\beta(\mu, \infty)$ (that is, reluctant dynamics converges more slowly for them) than all $\mu$ with $\Delta(\mu) > 0$.

We also consider some other candidate mechanisms for the non-universality we observe.
We will demonstrate that merely the property of having \textbf{discrete or continuous support} for a distribution does not determine a universality class for $\beta(\mu, \infty)$ in the way that we conjecture the discrepancy does; in particular, there are pairs of discrete $\mu_1$ and $\mu_2$ with $\beta(\mu_1, \infty) \neq \beta(\mu_2, \infty)$.
Similarly, the \textbf{moment matching} conditions that appear frequently as conditions for universality in probability theory broadly (such as the matching of means and variances we invoked above, which amounts to asking that the first 2 moments of $\mu$ take some specific values) and random matrix theory in particular (where a common condition asks to match 4 moments of entry distributions of Wigner matrices; see, e.g., \cite{tao2011random}) do not have a direct relationship with $\beta(\mu, \infty)$.
We do find that the \textbf{sparsity} of $\mu$ clearly affects $\beta(\mu, \infty)$ among the distributions of discrepancy 0: ``sparsifying'' $\mu$ in a suitable sense increases $\beta(\mu, \infty)$ provided that $\Delta(\mu) = 0$, but has no effect (consistent with Conjecture~\ref{conj:delta-pos-univ}) when $\Delta(\mu) > 0$.

\subsection{Organization}

The remainder of the paper is organized as follows.
In Section~\ref{sec:description-experiments} we describe the experiments we performed to estimate $\beta(\mu, \lambda)$ and the cases of $\mu$ we considered. 
In Section~\ref{sec:results1} we give our main results in support of the above Conjectures and other observations.
In Section~\ref{sec:results2} we give the results of some ancillary experiments and related speculations about the mechanisms by which non-universality might arise by analogy with well-understood behaviors in extreme value theory.

\section{Description of Experiments}
\label{sec:description-experiments}

\subsection{Estimating Runtime Scaling Laws}
\label{sec:estimate-scaling}

We now describe the kind of empirical evidence we will give for our Conjectures~\ref{conj:greedy-univ}, \ref{conj:delta-pos-univ}, and~\ref{conj:delta-zero-nonuniv}.
This will involve numerical estimates of $\beta(\mu, \lambda)$ produced for various choices of $\mu$ and $\lambda$.
We call such estimates $\what{\beta}(\mu, \lambda)$ and we describe below how we obtain these.

Given $\mu$ and $\lambda$, we fix a system size $N$, and generate some number $M$ of i.i.d.\ coupling matrices $J_{N, 1}, \dots, J_{N, M}$ with entry distribution $\mu$.
We always take $M = \num{1000}$ in our experiments unless stated otherwise.
Using the given $\lambda$, we run the $\lambda$-reluctant dynamics on each $J_{N, i}$ until it converges, and record the number of iterations $T_{N, i}$ required for this convergence.
We repeat this entire procedure for a sequence of values of $N$, say some $N_1 < \cdots < N_r$, giving a collection of results $(T_{N_a, i})_{a \in [r], i \in [M]}$.
For all of our results estimating the $\beta$ exponent, we consider the specific sequence $N \in \{25, 40, 50, 100, 150, 200, 300\}$.

We then perform linear regression on the collection of points $(\log N_a, \log T_{N_a, i})$.
We call $\what{\beta}$ the estimated slope and $\log \what{\alpha}$ the estimated intercept of this linear regression.
These give our estimated power law,
\[ \log T(N, \mu, \lambda) \approx \what{\beta} \log N + \log \what{\alpha}, \]
matching form of our conjectured scaling in \eqref{eq:log-power-law}.
In Section~\ref{sec:results1-scaling} we give some results similar to those of \cite{contucci2005interpolating} demonstrating, in support of Conjecture~\ref{conj:power-law}, that these scaling power laws hold, and also that our procedure achieves an excellent fit to empirical runtimes and a stable low-variance estimate of $\what{\beta}(\mu, \lambda)$ (relative to the much larger differences in these values for various $\mu$ that we will argue for in the rest of our experiments).
To illustrate these fluctuations in our estimates of $\beta$ (from the randomness in our sampled $J_{N, i}$), we often repeat the above procedure some number $K$ times for independent draws of the $J_{N, i}$, leading to several independent estimates $\what{\beta}_1, \dots, \what{\beta}_K$ and plot error bars calculated from this collection when we plot $\what{\beta}(\mu, \lambda)$ versus $\lambda$.

\begin{sidewaystable}
    \begin{center}
    \begin{tabular}{ccccccc}
        \hline
        $\mu$ & $\what{\beta}(\mu, 0)$ & $\what{\beta}(\mu, \infty)$ & $\Delta (\mu)$ & $\supp(\mu)$ & $\PP(J_{ij} = 0)$ & \makecell[c]{Moments 
    \\ Matching \\ $\sN(0, 1)$} \\
        \hline
        $\mathcal{N}(0, 1)$ & $\textbf{1.087} \pm 0.024$ & $\textbf{2.089} \pm 0.028$ & 0 & $\mathbb{R}$ & 0 & All \\
        $\mathrm{Unif}([-\sqrt{3}, \sqrt{3}])$ & $\textbf{1.087} \pm 0.023$ & $\textbf{2.077} \pm 0.023$ & 0 & $[-\sqrt{3}, \sqrt{3}]$ & 0 & 3\\
        Laplace & $\textbf{1.087} \pm 0.021$ & $\textbf{2.123} \pm 0.026$ & 0 & $\mathbb{R}$ & 0 & 3\\
        Student's $t$ & $\textbf{1.086} \pm 0.019$ & $\textbf{2.106} \pm 0.026$ & 0 & $\mathbb{R}$ & 0 & 3\\
        Hyperbolic Secant & $\textbf{1.079} \pm 0.023$ & $\textbf{2.108} \pm 0.025$ & 0 & $\mathbb{R}$ & 0 & 3\\
        \hline
        $\nu_1$ & $\textbf{1.094} \pm 0.026$ & $\textbf{1.990} \pm 0.022$ & 0 & $\{-1, 0, \sqrt{2}\}$ & $1 - \sqrt{2}/2$ & 3\\
        $\nu_2$ & $\textbf{1.099} \pm 0.025$ & $\textbf{1.923} \pm 0.025$ & 0 & $\{0, \pm 1, \pm \sqrt{6}\}$ & $1/3$ & 7 \\
        $\nu_3$ & $\textbf{1.101} \pm 0.025$ & $\textbf{1.672} \pm 0.024$ & 1 & $\{0, \pm 1, \pm 2, \pm3\}$ & $14/36$& 7 \\
        $\nu_4$ & $\textbf{1.099} \pm 0.020$ & $\textbf{1.667} \pm 0.024$ & 1 & $\{0, \pm 1, \pm 2, \pm3, \pm4\}$ & 115/288 & 9 \\
        \hline
        Rademacher & $\textbf{1.114} \pm 0.021$ & $\textbf{1.663} \pm 0.022$ & 1 & $\{-1, 1\}$ & 0 & 3\\
        $\Sparsify(\Rad, p = 1/3)$ & $\textbf{1.126} \pm 0.020$ & $\textbf{1.652} \pm 0.025$ & $\sqrt{3}$ & $\{- \sqrt{3}, 0, \sqrt{3}\}$ & $2/3$ & 5 \\ 
        $\Sparsify(\Rad, p = 1/5)$ & $\textbf{1.165} \pm 0.024$ & $\textbf{1.651} \pm 0.026$ & $\sqrt{5}$ & $\{- \sqrt{5}, 0, \sqrt{5}\}$ & $4/5$ & 3\\
        \hline
    \end{tabular}
    \end{center}
    \caption{\textbf{Summary of power law estimates and distribution properties.} For all of the distributions $\mu$ defined in Section~\ref{sec:dist} and discussed throughout the paper, we present estimates of $\what{\beta}(\mu, 0)$ and $\what{\beta}(\mu, \infty)$ computed as discussed in Section~\ref{sec:estimate-scaling}, as well as the discrepancy, support, sparsity (probability of taking the value 0), and number of moments matching the standard Gaussian distribution.}
    \label{table:big}
\end{sidewaystable}

\subsection{Choice of Distributions}
\label{sec:dist}

We now describe the various choices of probability measures $\mu$ that we use in our experiments.
Per the assumptions above, we always choose $\mu$ to have mean 0 and variance 1.

We consider the following well-known continuous distributions, whose densities we give together with the names we use:
\begin{align*}
    \text{Gaussian:} \,\,\,&\,\,\, \frac{1}{\sqrt{2\pi}} \exp\left(-\frac{1}{2}x^2\right)\,dx, \\
    \text{Uniform:} \,\,\,&\,\,\, \frac{1}{2\sqrt{3}}\bm 1\{x \in [-\sqrt{3}, \sqrt{3}]\}\,dx, \\
    \text{Laplace:} \,\,\,&\,\,\, \frac{1}{\sqrt{2}} \exp(-\sqrt{2}\,|x|)\,dx, \\
    \text{Hyperbolic Secant:} \,\,\,&\,\,\, \frac{1}{2}\mathrm{sech}\left(\frac{\pi}{2}x\right)\,dx, \\
    \text{Student's $t$:} \,\,\,&\,\,\, \frac{\Gamma(\frac{\nu + 1}{2})}{\sqrt{\pi(\nu - 2)} \,\Gamma(\frac{\nu}{2})}\left(1 + \frac{x^2}{\nu - 2}\right)^{-\frac{\nu + 1}{2}}\,dx \text{ for } \nu > 2.
\end{align*}
For the final distribution, in all experiments involving it we take the value $\nu = 6$.

For discrete distributions, we often work with the Rademacher distribution, which refers to the distribution
\[ \text{Rad} = \text{Rademacher} = \Unif(\{-1, +1\}) = \frac{1}{2}\delta_{-1} + \frac{1}{2}\delta_{+1}. \]
In addition, we will discuss the following somewhat ad hoc examples, which have various combinations of properties involving the discrepancy and number of moments agreeing with $\sN(0, 1)$ that we will make use of:
\begin{align*}
    \nu_{1} &= (\sqrt{2} - 1)\, \delta_{-1} +  \left(1 - \frac{\sqrt{2}}{2}\right)\, \delta_{0} + \left(1 - \frac{\sqrt{2}}{2}\right)\, \delta_{\sqrt{2}}, \\
    \nu_2 &= \frac{1}{30} \delta_{-\sqrt{6}} + \frac{3}{10} \delta_{-1} + \frac{1}{3}\delta_0 + \frac{3}{10}\delta_{1} + \frac{1}{30}\delta_{\sqrt{6}}, \\
    \nu_3 &= \frac{1}{180} \delta_{-3} + \frac{1}{20} \delta_{-2} + \frac{1}{4} \delta_{-1} + \frac{14}{36} \delta_0 + \frac{1}{4} \delta_{1} + \frac{1}{20} \delta_{2} + \frac{1}{180} \delta_{3}, \\
    \nu_4 &= \frac{1}{6720} \delta_{-4} + \frac{11}{2520} \delta_{-3} + \frac{13}{240} \delta_{-2} + \frac{29}{120} \delta_{-1} + \frac{115}{288} \delta_0 + \frac{29}{120} \delta_{1} + \frac{13}{240} \delta_{2} + \frac{11}{2520} \delta_{3}+ \frac{1}{6720}\delta_{4}.
\end{align*}

Finally, for both continuous and discrete distributions, we consider the following ``sparsification'' operation.
\begin{definition}
    \label{def:sparsify}
    Let $\mu$ be a probability measure on $\mathbb{R}$ with $\mathbb{E}_{X \sim \mu} X = 0$ and $\mathbb{E}_{X \sim \mu}X^2 = 1$, and let $p \in (0, 1)$.
    Let $\mathrm{Sparsify}(\mu, p)$ be the probability measure that we sample from by returning $\frac{1}{\sqrt{p}}X$ for $X \sim \mu$ with probability $p$ and zero with probability $1 - p$.
    Note that the first two moments of $\mathrm{Sparsify}(\mu, p)$ are the same as those of $\mu$.
\end{definition}

Some of the properties of all of the various distributions we work with that will be most important for our purposes are summarized in Table~\ref{table:big}, which the reader might find useful to consult periodically while reading the results below.

\section{Main Numerical Results: Runtime Scaling}
\label{sec:results1}

\subsection{Evidence for Scaling Laws of Conjecture~\ref{conj:power-law}}
\label{sec:results1-scaling}
 
\begin{figure}
    \centering 
    \includegraphics[width=1.0\textwidth]{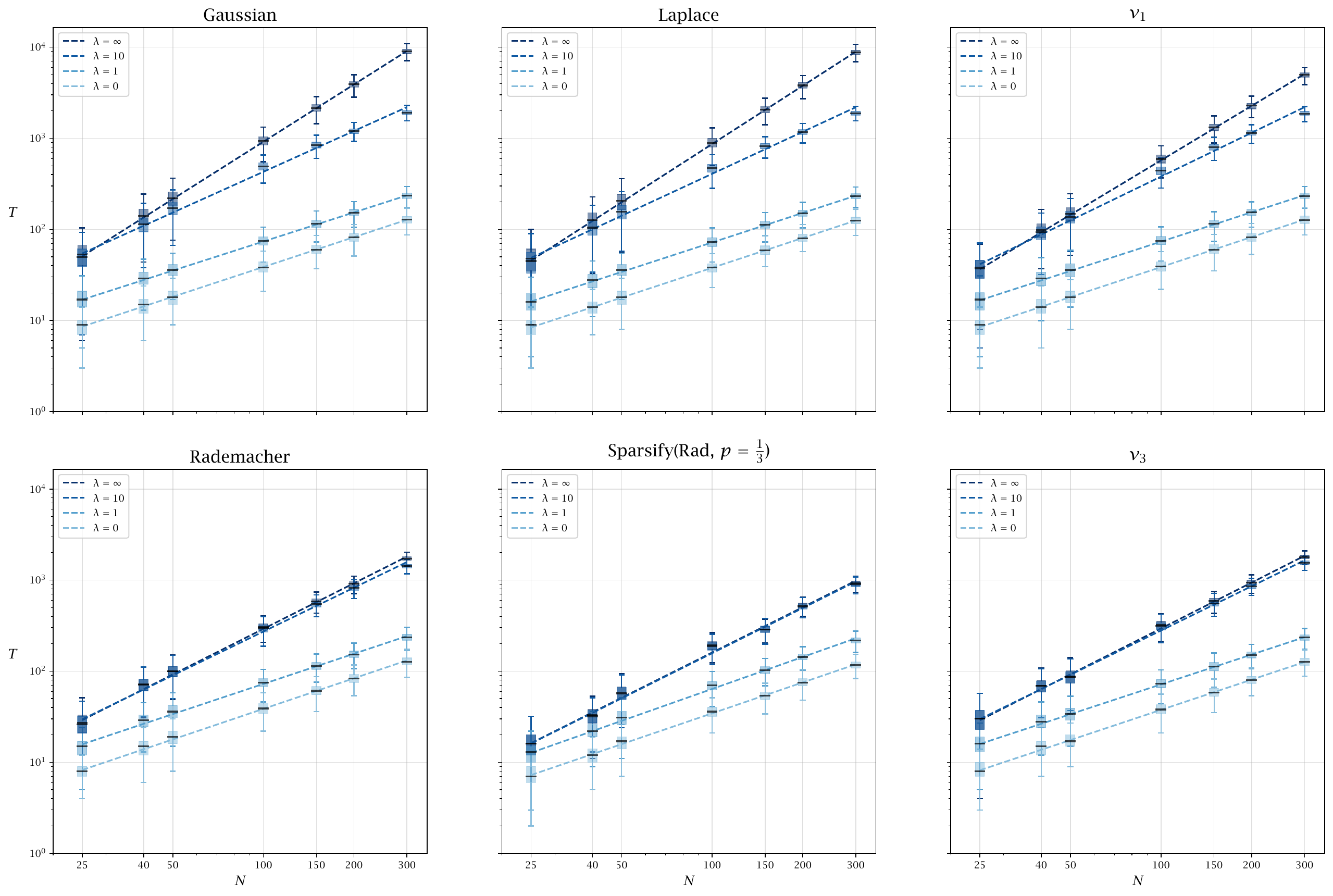} 
    \caption{\textbf{Illustration of power law fits.} For various choices of $\mu$ and $\lambda$, we plot power laws fit to the runtime $T$ as a function of $N$. Dotted lines show the fitted power law on a log-log plot, and box plots show the distribution of several random draws of the runtime over 50 trials for various $N$. Darker colors indicate larger \(\lambda\) (more reluctant dynamics).} 
    \label{fig:alpha_lam_scaling_general} 
\end{figure}

Figure~\ref{fig:alpha_lam_scaling_general} shows that our procedure gives qualitatively good fits of the empirical runtime data we gather, as described above.
In particular, it plots the results of a single linear regression as described above for various choices of $\mu$ and and $\lambda$.
In Table~\ref{table:variation_over_fits} we also report the minimum $R^2$ coefficient in any linear regression performed on all $\lambda$ considered for several choices of $\mu$, which are all at least \num{0.97}.

\begin{figure}
    \centering 
    \includegraphics[width=1.0\textwidth]{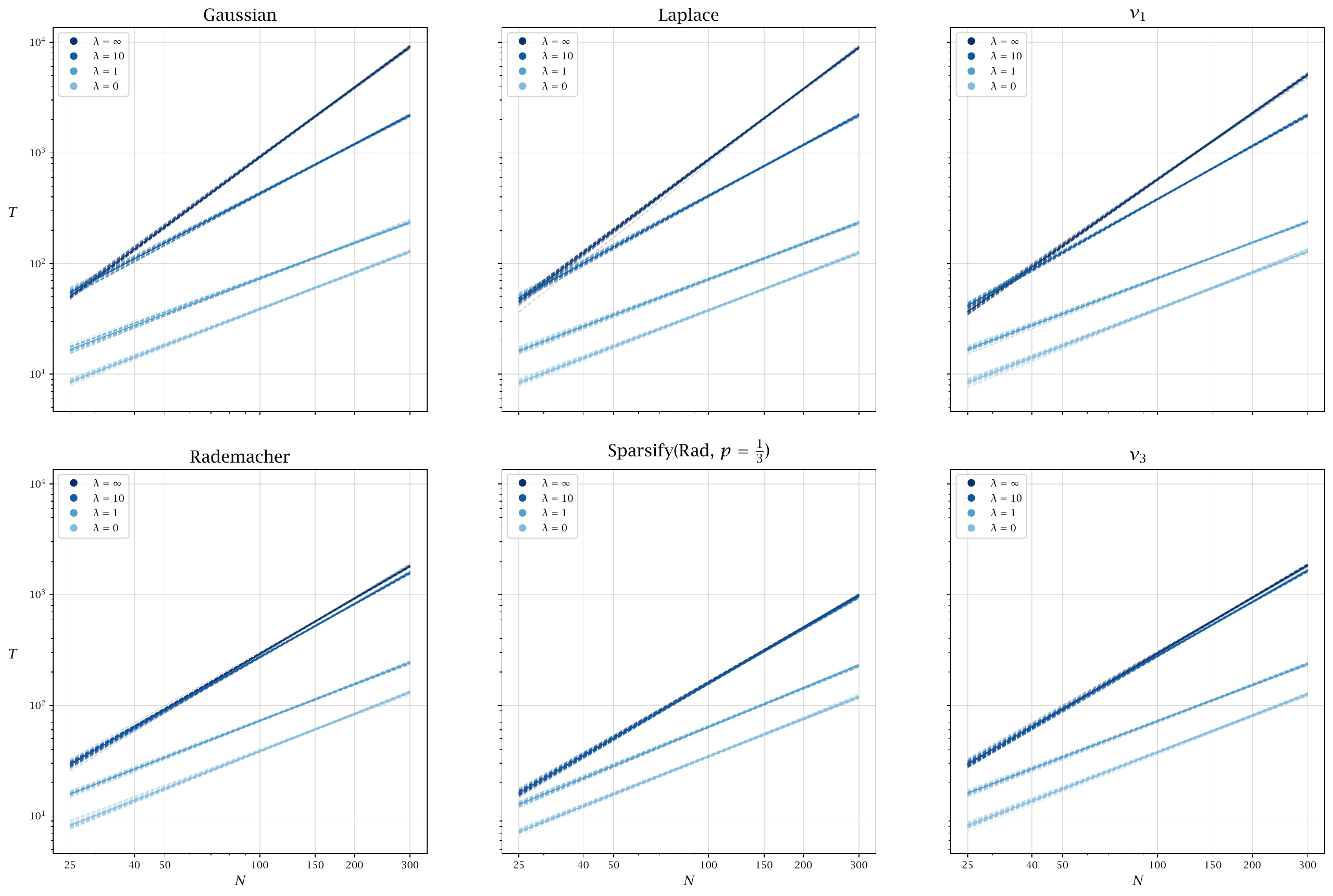} 
    \caption{\textbf{Fluctuations of power law fits.} For various $\mu$ and $\lambda$, we plot several linear fits on a log-log plot of the relationship between $T$ and $N$, estimated from independent experiments. The lines are close to one another for each such choice, and the thickness of the resulting ``bands'' reflects the fluctuations in our estimates of $\what{\beta}$ due to random sampling of the coupling matrices.} 
    \label{fig:many-lines} 
\end{figure}

\begin{figure}
    \vspace{2em}
    \centering 
    \includegraphics[width=\textwidth]{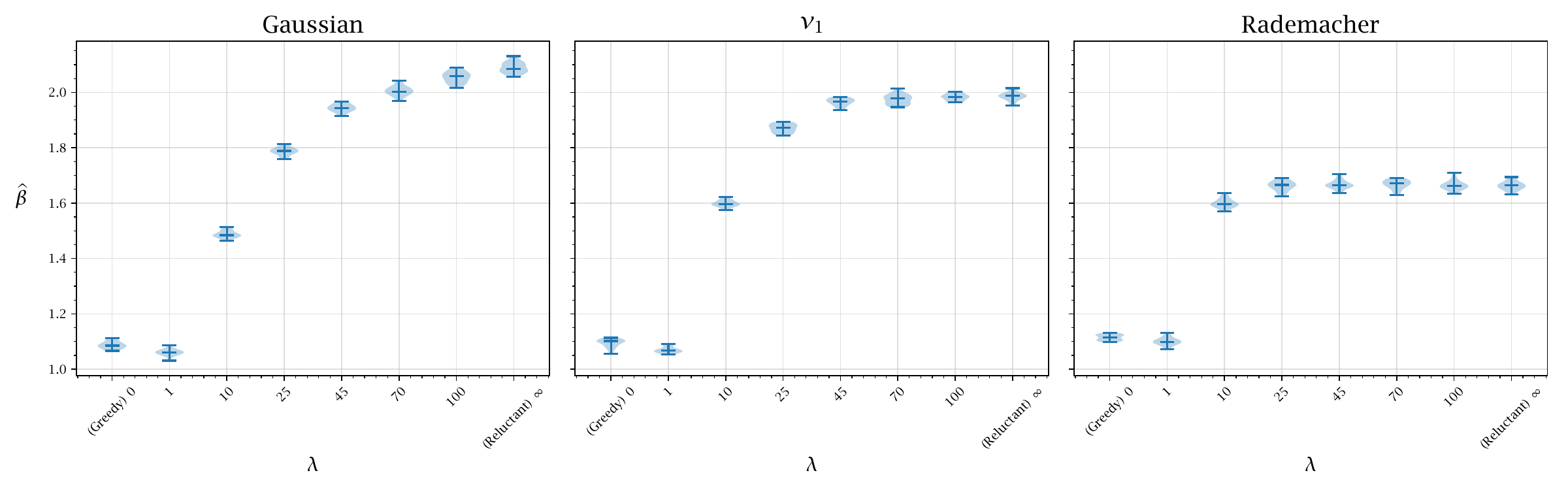} 
    \caption{\textbf{Fluctuations of exponent estimates.} We plot violin plot representations of the distribution of $\hat{\beta}(\mu, \lambda)$ estimated 50 repeated regressions using 20 independent draws of $J$ for each value of $N$, for various $\mu$ and $\lambda$.} 
    \label{fig:violin_beta_spread} 
\end{figure}

To assess the robustness of these slope estimates with respect to the randomness of the sampling, we also evaluate the variability of $\hat{\beta}$ across repeated regressions on independently sampled collections of $J$, i.e., of independent repetitions of the same experiment obtaining different estimates of $\hat{\beta}$.
In Figure~\ref{fig:many-lines}, we plot the regression lines obtained from many such regressions for several choices of $\mu$ and $\lambda$, and observe that these lines are well-concentrated, meaning that our estimates of $\hat{\beta}$ in particular do not fluctuate very much.
In Table~\ref{table:variation_over_fits} we report the mean and standard deviation of 10 independent estimates of $\hat{\beta}$ obtained from regressions using only $M = \num{100}$ independent draws of $J$ for each $N$. 
The empirical standard deviations here are relatively small, on the order of magnitude of $10^{-2}$ compared to means of the order $10^0$. 
As a last visualization, Figure~\ref{fig:violin_beta_spread} shows the empirical distribution of $\what{\beta}(\mu, \lambda)$ obtained from $K = 50$ independent regressions using $M = 20$ independent draws of $J$ per value of $N$ in each regression. Even with substantially fewer points per fit, the distribution remains approximately symmetric and concentrated around the mean. 
It is also evident that the observed spread of the $\hat{\beta}$ estimates is often significantly smaller than the differences across distributions $\mu$ in several cases, for example between $\mu$ Gaussian and Rademacher.
Thus we do not expect the variability of our random estimates of $\hat{\beta}$ to substantially affect our conclusions about non-universality.

On the whole, we conclude from these experiments that our estimates $\what{\beta}(\mu, \lambda)$ are reliable, and also obtain evidence that Conjecture~\ref{conj:power-law} on the power law scaling of the runtime indeed holds.
Having established this, for the remainder of this section we focus on the behavior of our estimates of $\hat{\beta}$, without returning to the individual log-log plots underlying each such estimate.

\begin{table}
\centering
\small
\begin{tabular}{lccccc}
\hline \\[-1em]
$\mu$ (Distribution) & $\what{\beta}{(\mu, 0)}$ & $\what{\beta}{(\mu, \infty)}$ & Minimum $R^2$\\[0.25em]
\hline
Gaussian   & $\textbf{1.087} \pm 0.010$ & $\textbf{2.089} \pm 0.013$ & 0.973\\
Laplace    & $\textbf{1.087} \pm 0.011$ & $\textbf{2.123} \pm 0.013$ & 0.972\\
$\nu_1$ & $\textbf{1.094} \pm 0.010$ & $\textbf{1.990} \pm 0.010$ & 0.971\\
Rademacher & $\textbf{1.114} \pm 0.011$ & $\textbf{1.663} \pm 0.012$ & 0.972\\
Sparsify(Rad, p = 1/3) & $\textbf{1.126} \pm 0.010$ & $\textbf{1.652} \pm 0.014$ & 0.974\\
\hline
\end{tabular}
\caption{\textbf{Power law fits.} Estimated scaling exponents $\hat{\beta}(\mu,0)$ and $\hat{\beta}(\mu,\infty)$, each computed from 100 independent regressions (10 data points per regression). Reported standard deviations reflect the empirical variation of the slope estimates across those 100 regressions. Minimum \(R^2\) value (over all values of $\lambda$ considered) for quality of linear regression fit for several distributions.}
\label{table:variation_over_fits}
\end{table}

\subsection{Initial Evidence of Universality and Non-Universality}

\begin{figure}
    \centering 
\includegraphics[width=\textwidth]{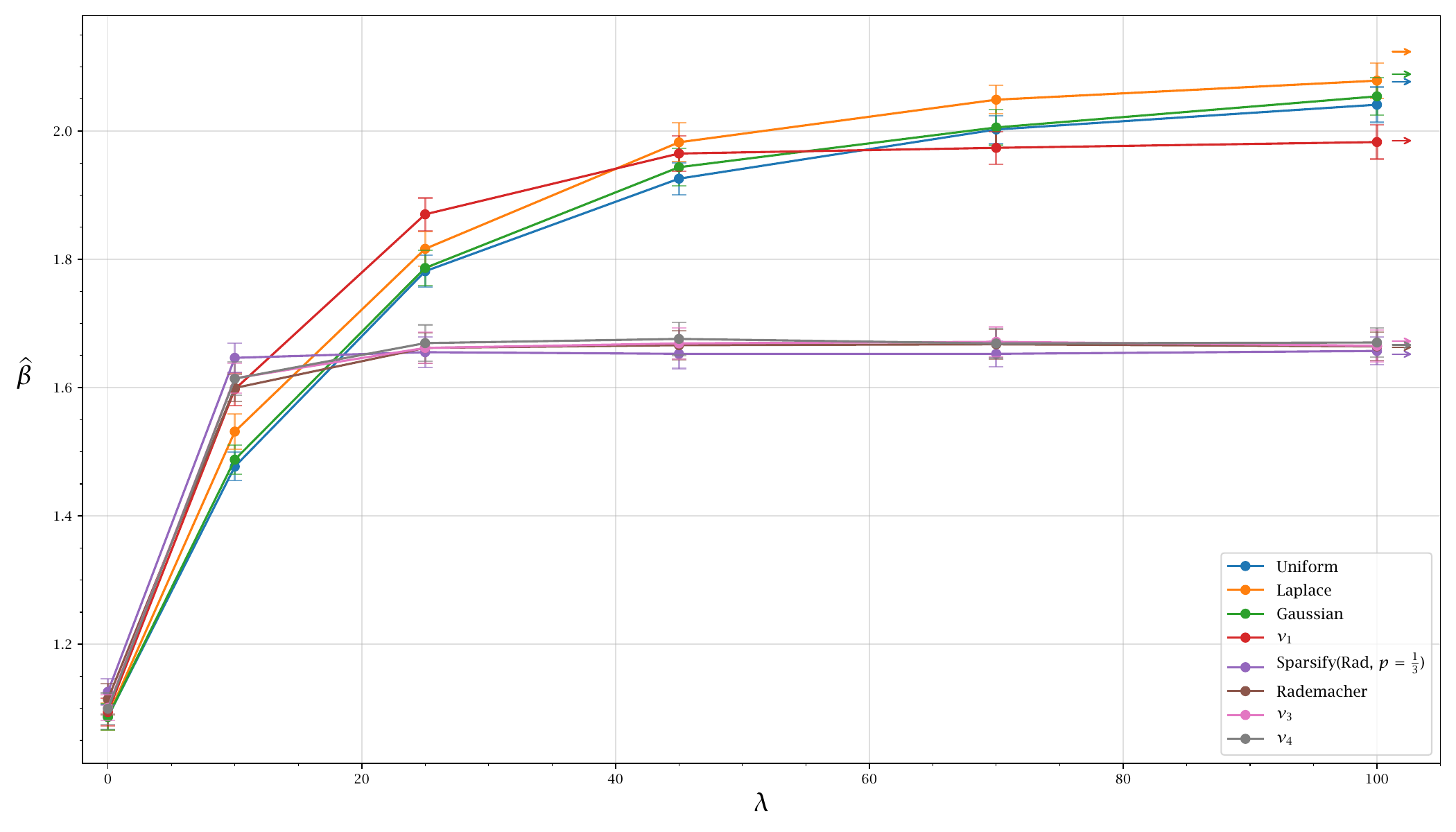} 
\caption{\textbf{Demonstration of non-universality.} For various entry distributions $\mu$, we plot $\what{\beta}(\mu, \lambda)$ over a range of $\lambda$ with error bars over several independent estimates of $\what{\beta}$. We clearly observe that for small $\lambda$ (greedy dynamics) the estimated values are similar, while for large $\lambda$ (reluctant dynamics) the estimated values are clearly different beyond errors due to our approximation procedure.} 
    \label{fig:scaling} 
\end{figure}

We may summarize all our results at the level of the estimates $\what{\beta}(\mu, \lambda)$ by plotting these estimates for various $\lambda$ for several choices of $\mu$.
We give one such plot in Figure~\ref{fig:scaling}, and discuss some of its features below.

At \(\lambda = 1\) (close to a greedy algorithm), estimated $\what{\beta}$ values cluster near \num{1.1} across several distributions, but by \(\lambda = 100\) and more clearly at $\lambda = \infty$, they separate into two classes: Gaussian, Laplace, and $\nu_1$ converge as $\lambda \to \infty$ to various considerably different values near $2$, while Rademacher and $\Sparsify(\Rad, p = 1/3)$ converge to what appears to be a single value near \num{1.66}.
This is consistent with the universality for greedy algorithms of Conjecture~\ref{conj:greedy-univ}, as well as the non-universality for reluctant algorithms of Conjecture~\ref{conj:delta-zero-nonuniv}.
We also notice that the latter class consists precisely of those $\mu$ (among the ones considered here) with $\Delta(\mu) > 0$, which have very similar values of $\what{\beta}(\mu, \lambda)$ for all $\lambda$, but especially for $\lambda = \infty$, while the former class is of those $\mu$ with $\Delta(\mu) = 0$, and these have substantially different values of $\what{\beta}(\mu, \infty)$.
This is also consistent with the full statements of Conjectures~\ref{conj:delta-pos-univ} and~\ref{conj:delta-zero-nonuniv}.
We report specific estimated numbers in Table~\ref{table:big} for $\lambda = 0$ and $\lambda = \infty$, which are also compatible with these claims.

\begin{figure}
    \centering 
\includegraphics[width=\textwidth]{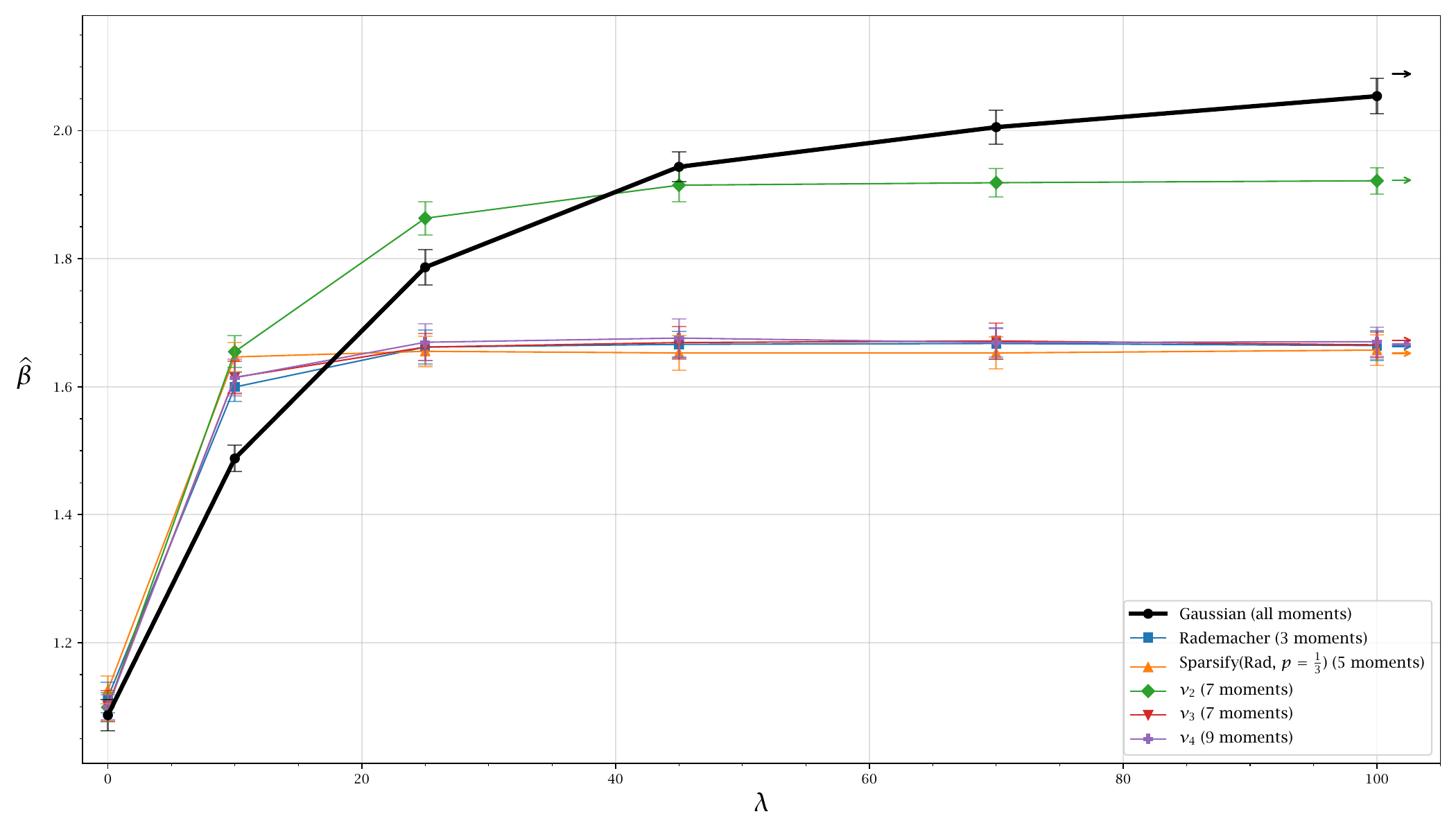} 
    \caption{\textbf{Moment matching to Gaussian.} For entry distributions $\mu$ matching various numbers of moments of the Gaussian distribution, we plot $\what{\beta}(\mu, \lambda)$ over a range of $\lambda$ with error bars over several independent estimates of $\what{\beta}$. The estimates for the Gaussian distribution are shown by the bold black line, and the numbers of matching moments for each distribution are given in the legend.} 
\label{fig:moment_matching_fig} 
\end{figure}

\subsection{Moment Matching}

We next consider a few other mechanisms other than discrepancy that might be intuitive candidates for explaining the behavior observed above.
One reasonable hypothesis is that such differences could be explained by the moments of distributions: the more moments that two distributions share, the more similar their scaling behavior might be expected to become. 
To test this, we consider distributions that match the Gaussian distribution $\sN(0, 1)$ up to various orders of moments.
We present the results in Figure~\ref{fig:moment_matching_fig}, which repeats some of the results from Figure~\ref{fig:scaling}, but considers a different family of distributions chosen to match different numbers of Gaussian moments. 

Although the Rademacher, $\Sparsify(\Rad, 1/3)$, $\nu_3$, and $\nu_4$ distributions successively match three, five, seven, and nine moments of the Gaussian distribution, the scaling exponents they approach under reluctant dynamics do not become progressively closer to the Gaussian ones as the number of moments matched grows. 
Instead, these four distributions exhibit nearly identical limiting behavior, consistent with Conjecture~\ref{conj:delta-pos-univ} since all of these choices of $\mu$ have positive discrepancy $\Delta(\mu) > 0$.

On the other hand, while $\nu_2$ matches seven moments of $\sN(0, 1)$, it is the only one of these distributions that lies closer to $\sN(0, 1)$ in terms of estimated $\what{\beta}$ values.
Since $\Delta(\nu_2) = 0$ unlike the above distributions, in this example the discrepancy of a distribution explains a distinction in estimated $\what{\beta}$ values, while moment matching does not.
Indeed, our results suggest that there is no direct relationship between the scaling exponents $\beta(\mu, \lambda)$ and the number of moments of $\sN(0, 1)$ that $\mu$ match.
Having refuted this proposal we have not considered moment comparisons with distributions other than $\sN(0, 1)$, but we expect a similar absence of any clear relationship.

\subsection{Discrete Versus Continuous}
Another reasonable hypothesis is that the support of \(\mu\) being discrete or continuous could explain the differences in runtime scaling behavior. Indeed, a valid interpretation of most but not all of the results in Figure~\ref{fig:scaling} is that, for the case of reluctant algorithms ($\lambda = \infty$), continuous distributions yield \(\what{\beta} (\mu, \infty) \gtrsim 2.0\) while discrete distributions have \(\what{\beta}(\mu, \infty) \approx 1.6\). 

\begin{figure}
    \centering 
    \includegraphics[width=\textwidth]{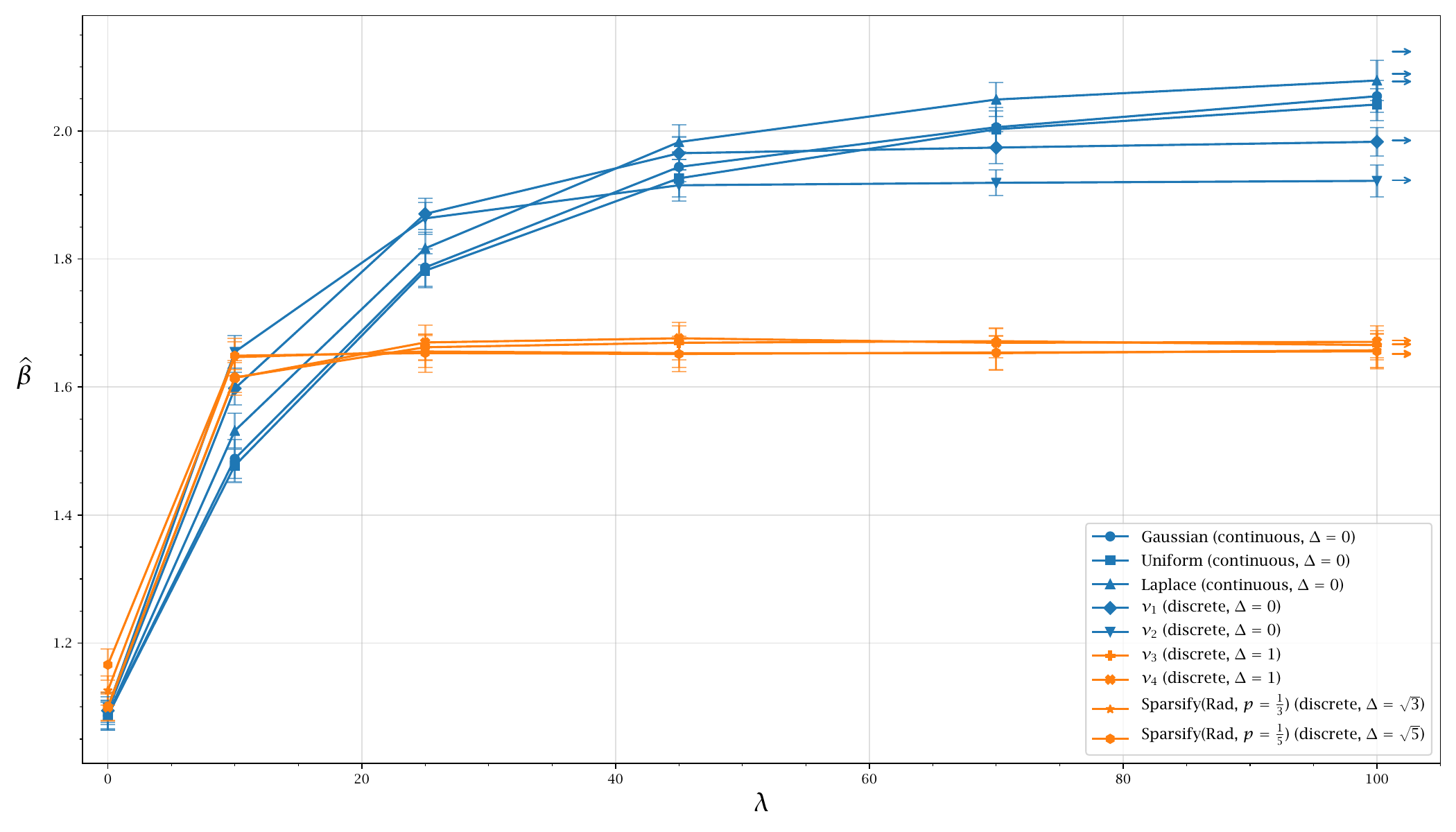} 
    \caption{\textbf{Discrete support, continuous support, and discrepancy.} For entry distributions $\mu$ with various combinations of discrete or continuous support and positive or zero discrepancy, we plot $\what{\beta}(\mu, \lambda)$ over a range of $\lambda$ with error bars over several independent estimates of $\what{\beta}$. The plots are colored according to whether the discrepancy of $\mu$ is zero or positive.} 
    \label{fig:discrete-vs-continuous} 
\end{figure}

However, the example of $\mu = \nu_1$ violates this pattern: this distribution has discrete support, but its support contains both a rational and an irrational point, so $\Delta(\nu_1) = 0$.
We recall its definition: $X \sim \nu_1$ is given by
\[
X = \begin{cases}
-1 & \text{with probability } \sqrt{2} - 1, \\
0  & \text{with probability } 1 - \frac{\sqrt{2}}{2},  \\
\sqrt{2} & \text{with probability } 1 - \frac{\sqrt{2}}{2}.
\end{cases} \]
Accordingly, $\what{\beta}(\nu_1, \infty) \approx \num{1.99}$ is much closer to the values estimated for continuous distributions than all of the other discrete distributions, which have $\Delta(\mu) > 0$.
The same also applies to $\nu_2$.
We show these results in Figure~\ref{fig:discrete-vs-continuous}.
Thus, again, the discrepancy of probability measures offers a much more convincing explanation of the variation in runtime scaling laws than the simpler property of discreteness of the support.

To investigate this effect from a different perspective, we also propose an interpolation between continuous and discrete distributions.
We consider a ``starting'' discrete distribution $\Rad = \Unif(\{-1, +1\})$ and consider a coupling matrix parametrized by $s \geq 0$ with entries
\[\sqrt{N} \cdot J_{ij}^{(s)} = e^{-s}X_{ij} + \sqrt{1 - e^{-2s}}\,Y_{ij} \]
where $X_{ij} \sim \Unif(\{\pm 1\})$ and $Y_{ij} \sim \sN(0, 1)$ are all drawn independently. 
Thus $J_{ij}^{(s)}$ always maintains our normalization of having mean 0 and variance 1, but becomes ``more continuous'' as $s \to \infty$.
(A probabilistic interpretation is that each entry of $J$ undergoes an independent Ornstein-Uhlenbeck process as $s$ grows; in random matrix theory such an operation on a matrix is known as \emph{Dyson Brownian motion}).
Let us also write $\mu^{(s)}$ for the entrywise distribution of such $J$ for a given $s \geq 0$.

\begin{figure}
    \centering \includegraphics[width=\textwidth]{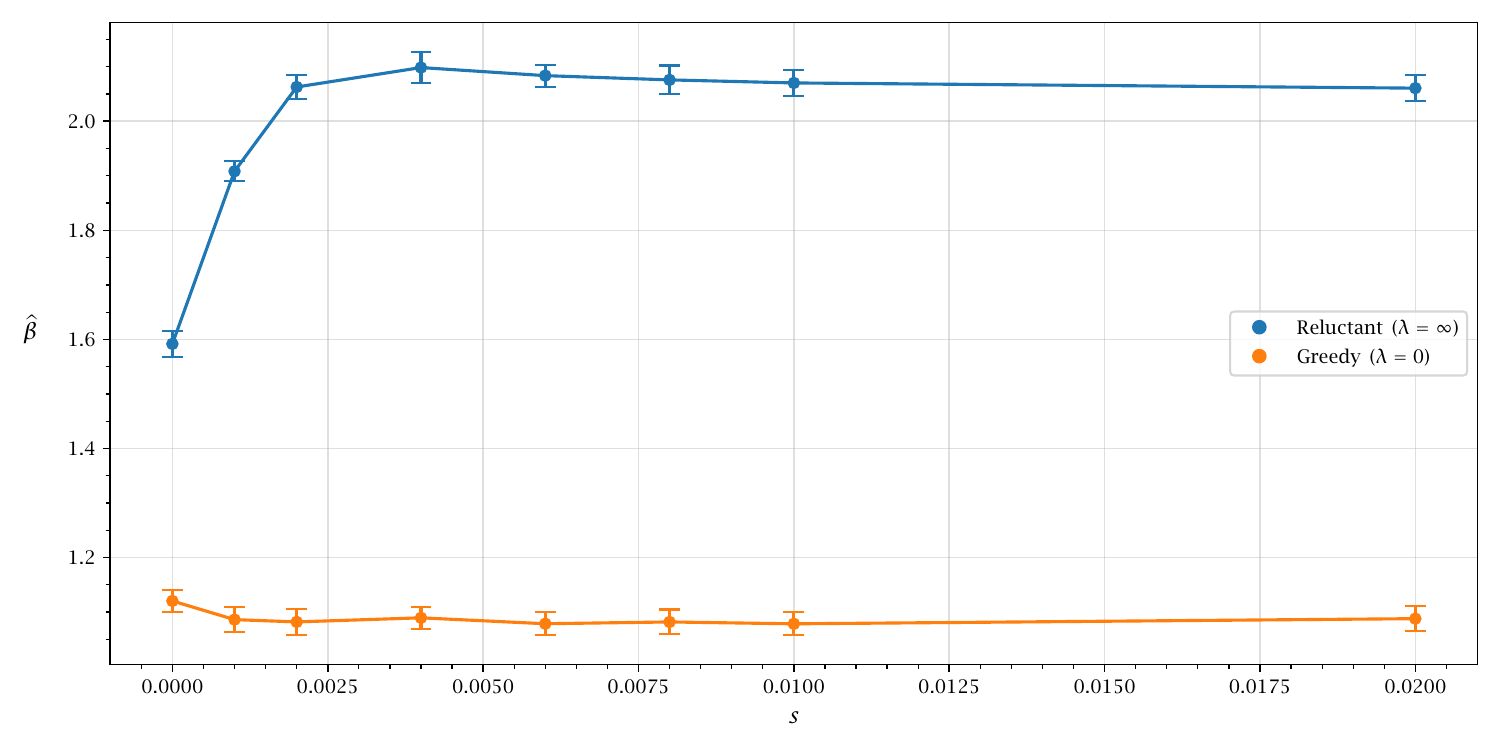} 
    \caption{\textbf{Runtime under entrywise Ornstein-Uhlenbeck dynamics.} We plot estimated values of $\what{\beta}$ for different entry distributions $\mu^{(s)}$ formed by running Ornstein-Uhlenbeck dynamics for an amount of time $s$ on an initial distribution $\mu^{(0)} = \mathrm{Unif}(\{\pm 1\})$.} 
    \label{fig:gaussian noise} 
\end{figure}

We present the runtime scalings for reluctant and greedy dynamics for such $J = J^{(s)}$ for various $s$ in Figure~\ref{fig:gaussian noise}.
We note that $\Delta(\mu^{(0)}) > 0$ while $\Delta(\mu^{(s)}) = 0$ for all $s > 0$.
And indeed, we observe that $\what{\beta}(\mu^{(s)}, \infty)$ (the estimated runtime scaling for reluctant dynamics) quickly grows as $s$ increases and reaches values close to those estimated for $\mu = \sN(0, 1)$ above.
On the other hand, $\what{\beta}(\mu^{(s)}, 0)$ (the estimated runtime scaling for greedy dynamics) changes very little as $s$ grows.
Both observations are again compatible with Conjectures~\ref{conj:greedy-univ} and~\ref{conj:delta-zero-nonuniv}.

\subsection{Sparsity}

Finally, we consider the sparsity of $\mu$ as it relates to the scaling exponent $\beta(\mu, \lambda)$.
We consider sparsifying (using the operation from Definition~\ref{def:sparsify}) the Gaussian, Rademacher, and $\nu_1$ distributions, which respectively have continuous support, discrete support and positive discrepancy, and discrete support and zero discrepancy.
For each of these $\mu$, we plot $\what{\beta}(\Sparsify(\mu, p), \lambda)$ for $\lambda \in \{0, \infty\}$ and $p \in [0, 1]$ on an evenly spaced grid.

\begin{figure} 
    \centering 
    \includegraphics[width=\textwidth]{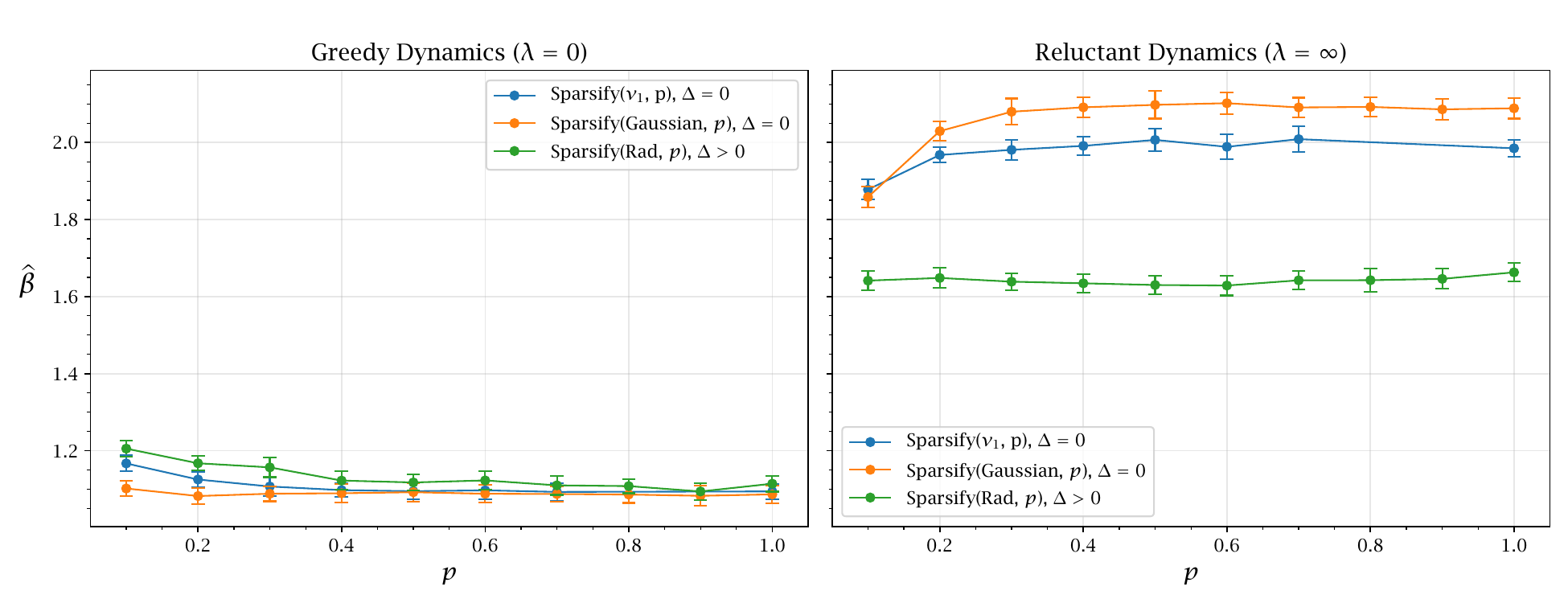} 
    \caption{\textbf{Sparsity.} We plot the estimated scaling exponent $\what{\beta}(\Sparsify(\mu, p))$ for various $\mu$ and greedy and reluctant dynamics.
    For each $(\mu, p)$, we plot the mean and standard deviation over $K = 50$ independent estimates of $\what{\beta}$, each computed using $M = 20$ independent samples of the coupling matrix for each value of $N$.} 
    \label{fig:example}
\end{figure}

We present the results in Figure~\ref{fig:example}.
These clearly demonstrate that the scaling exponent $\what{\beta}(\Sparsify(\mu, p), 0)$ of greedy dynamics does not depend on $p$, consistent with Conjecture~\ref{conj:greedy-univ}.
This is less clear for small values of $p$, but in this case as the coupling matrix becomes very sparse we suspect that various high-dimensional concentration of measure phenomena might not yet ``kick in.''
We leave a more careful investigation of this matter with larger-scale experiments to future work.

On the other hand, for reluctant dynamics, we see that $\what{\beta}(\Sparsify(\mu, p), \infty)$ has two different behaviors.
When $\Delta(\mu) = 0$, then this quantity increases with $p$, and converges to different numbers at $p = 1$ for $\mu = \sN(0, 1)$ and $\mu = \nu_1$.
On the other hand, when $\Delta(\mu) > 0$, then this quantity appears to be constant in $p$.
Both observations are again consistent with Conjectures~\ref{conj:delta-pos-univ} and~\ref{conj:delta-zero-nonuniv}, as well as our remark around the Conjectures that, for reluctant dynamics when $\Delta(\mu) = 0$, sparsifying $\mu$ appears to increase the runtime.

\section{Additional Results: Distributions of Energy Increments}
\label{sec:results2}

We next present some calculations suggesting a possible explanation for why the discrepancy of a measure is so closely related to the behavior of reluctant dynamics on coupling matrices with that distribution of entries.
Namely, we consider the distribution of the size of the first step taken by greedy and reluctant dynamics from a uniformly random initialization.

\subsection{Heuristic Derivation of Distribution}

We consider the distribution of this first step from an initial position $\sigma \sim \Unif(\{\pm 1\}^N)$.
Suppose that $\sigma$ and $\sigma^{\prime}$ differ in a single coordinate $a$, so that $\sigma_a = -\sigma_a^{\prime}$ while $\sigma_i = \sigma_i^{\prime}$ for all $i \neq a$.
Then, we have
\[ \delta(J, \sigma, a) \colonequals H(J, \sigma^{\prime}) - H(J, \sigma) = \frac{1}{2}\sum_{i, j = 1}^N J_{ij} (\sigma_i\sigma_j - \sigma^{\prime}_{i} \sigma^{\prime}_{j}) = \sigma_a \sum_{i \in [N] \setminus\{a\}} 2J_{ai} \sigma_i. \]
When we take $\sigma \sim \mathrm{Unif}(\{\pm 1\}^N)$ and $J_{ij}$ to have mean 0 and variance $1 / N$ as in our normalization throughout, then the inner sum is a sum of $N - 1$ i.i.d.\ random variables of mean zero and variance $\EE (J_{ai} \sigma_i)^2 = 1 / N$.
By the central limit theorem, we expect such a sum to have law close to 
\[ \mathrm{Law}(\delta(J, \sigma, a)) \approx \mathcal{N}\left(0, (N - 1) \cdot \frac{4}{N}\right) \approx \mathcal{N}(0, 4). \]
(Multiplying this random variable by the independent random sign $\sigma_a$ will not change its law.)

This gives a reasonable prediction for the distribution of the change in energy associated to flipping any given spin.
One may also compute the covariance structure of these changes, as we have, for $a \neq b$,
\[ \EE \delta(J, \sigma, a) \delta(J, \sigma, b) = 4\EE \sigma_a \sigma_b \sum_{i, j = 1}^N J_{ai} J_{bj} \sigma_i \sigma_j = 4\sum_{i, j = 1}^N \EE[J_{ai} J_{bj}] \EE[\sigma_a\sigma_b\sigma_i\sigma_j] = \frac{4}{N}, \]
since when $a \neq b$ the expectation over $J$ can only be non-zero if $j = a$ and $b = i$.
To simplify our heuristic calculations, we assume below that this positive covariance is negligible.
Thus, our final heuristic prediction is that the vector $\delta(J, \sigma) = (\delta(J, \sigma, a))_{a = 1}^N$ has law approximately $\mathcal{N}(0, 4 I_N)$.

\subsection{Sizes of First Steps}

By definition, in the above setting, the first step taken by greedy dynamics will change the objective by the most negative value of the $\delta(J, \sigma)$, while the first step taken by reluctant dynamics will change the objective by the least negative (but still negative) of the same values.
Let us consider the universality of these quantities, which we call $\delta_{\greedy}(J, \sigma)$ and $\delta_{\reluctant}(J, \sigma)$, respectively.

These questions are most clearly understood through the lens of extreme value theory (we will only require the basic result of Exercise~3.2.2 of \cite{Durrett-1995-ProbabilityTheoryExamples}, but see also \cite{de2006extreme} for a general reference).
If indeed the vector $\delta(J, \sigma)$ can be approximated to have distribution $\mathcal{N}(0, 4I_N)$, then the Fisher-Tippett-Gnedenko theorem (the result cited above) implies that $\delta_{\greedy}(J, \sigma)$, viewed as the negative of the maximum of several i.i.d.\ random variables, asymptotically has a Gumbel distribution, while $\delta_{\reluctant}(J, \sigma)$, viewed as the maximum of several i.i.d.\ random variables with an absolutely continuous distribution that is bounded above, asymptotically has an exponential distribution.

\begin{figure}
    \centering 
    \includegraphics[width=\textwidth]{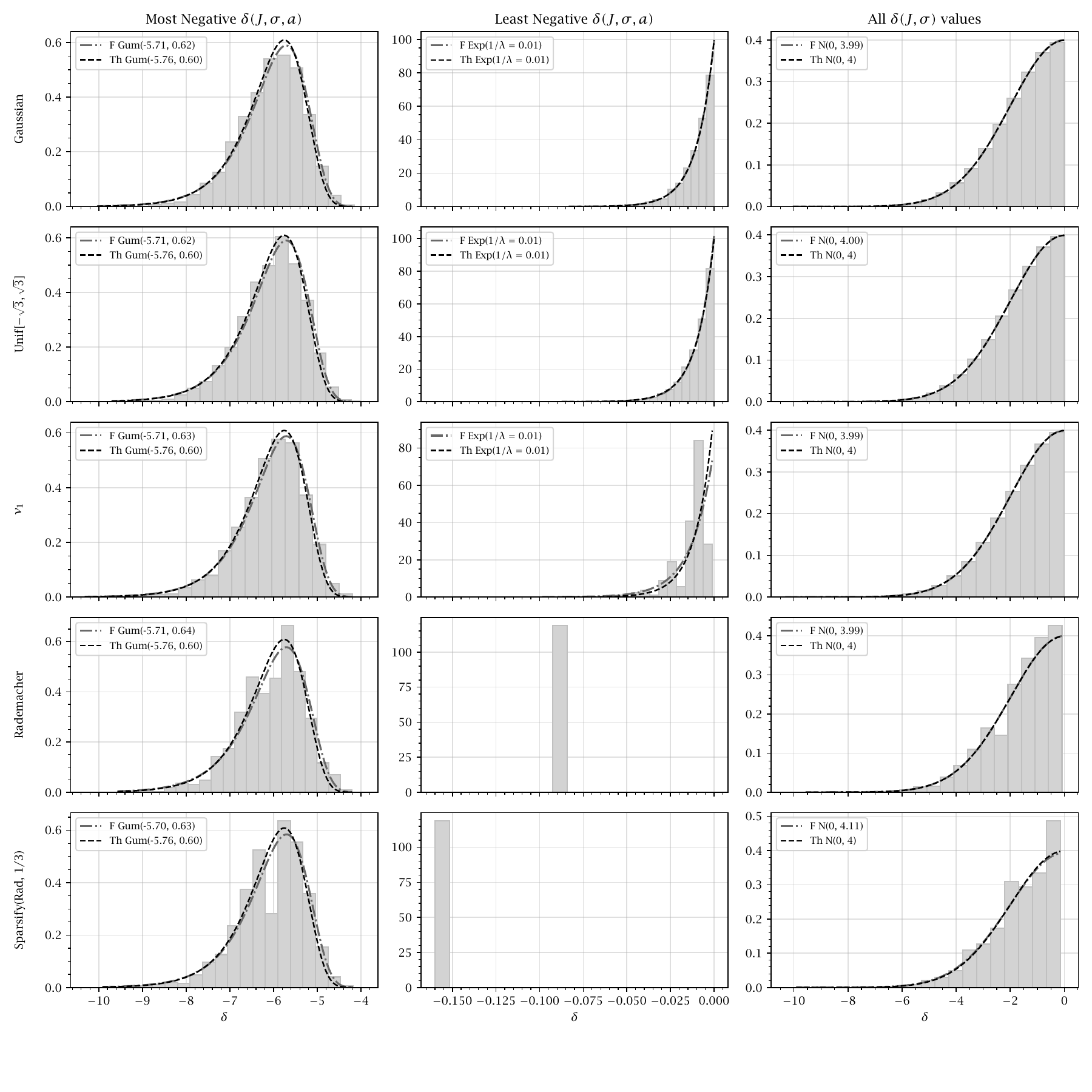} 
    \caption{\textbf{Distribution of step sizes.} For several choices of $\mu$, we plot the most negative, least negative, and all values of the vector of step sizes $\delta(J, \sigma)$ over several independent draws of $J$ and $\sigma$ drawn uniformly at random from $\{\pm 1\}^N$, along with approximate distributions from various families according to our theoretical predictions.}
    \label{fig:deltaE} 
\end{figure}

We observe in Figure~\ref{fig:deltaE} that these claims indeed are sound always for $\delta_{\greedy}(J, \sigma)$, and for distributions of zero discrepancy for $\delta_{\reluctant}(J, \sigma)$.
We consider Gaussian, $\Unif([-\sqrt{3}, \sqrt{3}])$,  $\nu_1$, Rademacher, and $\Sparsify(\Rad, p = 1/3)$ coupling distributions, generate 5000 coupling matrices $J$ and random initializations $\sigma \sim \Unif(\{\pm 1\}^N)$ for each distribution in dimension $N = 500$, and plot histograms of all entries in $\delta(J, \sigma)$ as well as of the values of $\delta_{\greedy}(J, \sigma)$ and $\delta_{\reluctant}(J, \sigma)$.

We consider both fitting densities from the Gumbel and exponential classes (marked ``F'' in the Figure) and plotting densities with parameters as predicted by theoretical calculations (marked ``Th''); as expected, these two approaches agree closely.
The theoretical settings of parameters are as follows.
For the Gumbel distribution, for a given $N$, as we are considering Gaussian random variables with variance $4 = 2^2$, we expect the distribution to have location and scale parameters $-2(\Phi^{-1}(1 - 1/N))$ and $2(\Phi^{-1}(1 - 1/eN) - \Phi^{-1}(1 - 1/N))$, respectively, where $\Phi$ is the cumulative distribution function of the standard Gaussian distribution $\sN(0, 1)$.
For the exponential distribution, note that the maximum $M$ of the negative values among $\delta \sim \sN(0, 4I_N)$ has roughly, conditioning on the signs of the $\delta_i$,
\begin{align*}
\PP\left[M > -\frac{t}{N}\right] 
&\approx \left(1 - \PP\left[|\delta_i| > \frac{t}{N}\right]\right)^{N / 2} \\
&\approx \left(1 - \frac{2t}{N} \cdot \frac{1}{\sqrt{8\pi}}\right)^{N / 2} \\
&\approx \exp\left(-\frac{t}{\sqrt{8\pi}}\right),
\end{align*}
so we expect an exponential distribution with rate parameter $N / \sqrt{8\pi}$.

On the other hand, we also observe that $\delta_{\reluctant}(J, \sigma)$ does \emph{not} obey an exponential distribution for distributions of positive discrepancy (the latter two listed above, Rademacher and its sparsified version); indeed, in these cases we observe that $\delta_{\reluctant}(J, \sigma)$ takes on the same value in nearly all random trials.
The explanation for this is that $\delta_{\reluctant}(J, \sigma)$ itself has a discrete distribution, since it can be no less negative than $-\Delta(\mu) / \sqrt{N}$ when the entry distribution is $\mu$.
A straightforward combinatorial calculation shows that, at least for simple entry distributions like $\mu = \Rad$ or $\mu = \Sparsify(\Rad, p)$, in fact with high probability some spin flip will achieve this minimum possible discrepancy.
More broadly, the issue is that we cannot apply the central limit theorem so naively as we did above to understand reluctant dynamics: since its behavior depends on the values of $\delta(J, \sigma)$ near zero, we would instead need to appeal to a \emph{local central limit theorem}.
But, these results do not apply (in a strong form) to distributions of positive discrepancy, for precisely the reason we have encountered (see, for instance, discussion of local limit theorems in Section~3.5 of \cite{Durrett-1995-ProbabilityTheoryExamples}).

\section{Additional Results: Final Energy Levels}

It is reasonable to wonder whether not only the runtime but the actual performance of reluctant dynamics is non-universal in terms of the final energy achieved after the algorithm converges.
This would be quite surprising, since the asymptotic ground state energy itself is known to be universal across broad classes of coupling distributions \cite{carmona2004universalitysherringtonkirkpatricksspinglass}, and at least for Gaussian couplings reluctant dynamics are believed according to the most detailed numerical studies to (nearly) reach that ground state energy \cite{EBKZ-2024-QuenchesSKReluctant}.
Thus, if reluctant dynamics performed worse for other distributions, it would also not share the qualitative property of (nearly) reaching the ground state energy, which seems unlikely.

\begin{figure}
    \centering 
    \includegraphics[width=\textwidth]{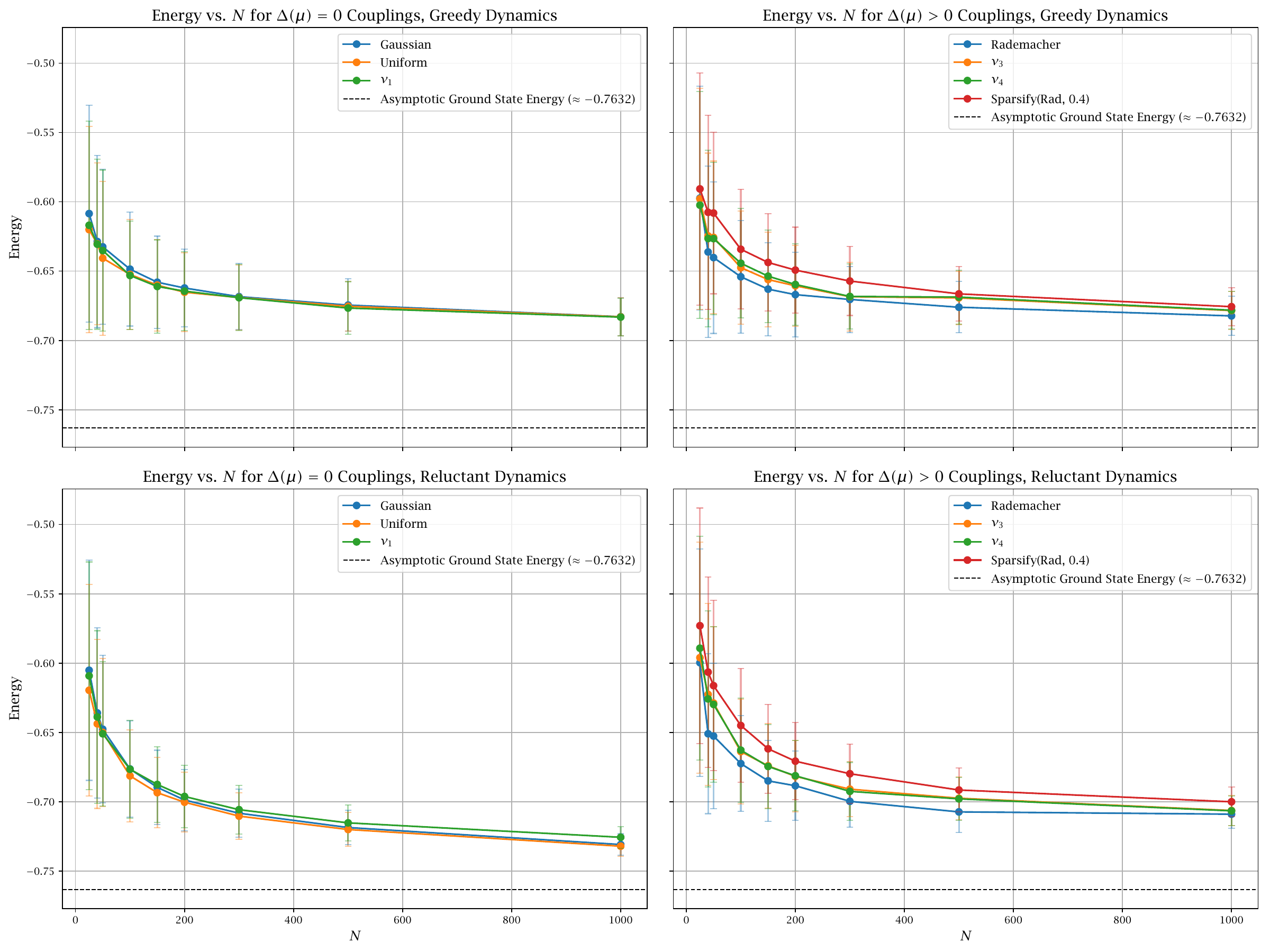} 
    \caption{\textbf{Final energy level at convergence.} We compare the energy level achieved at convergence by greedy and reluctant algorithms for both zero and positive-discrepancy distributions of couplings. Dashed lines in all plots indicate the theoretical asymptotic ground state energy as $N \to \infty$.}
    \label{fig:energy_vs_n} 
\end{figure}

In Figure~\ref{fig:energy_vs_n}, we present numerical experiments on the same scale as those substantiating our other predictions for the energy reached by greedy and reluctant dynamics for various coupling distributions, as a function of the system size $N$.
The results are inconclusive: the difference in final energy level between different distributions, particularly between distributions with zero discrepancy and those with positive discrepancy, is greater for reluctant dynamics than greedy dynamics, but also reluctant dynamics is still quite far from reaching the true asymptotic ground state energy for the system sizes we consider, and thus remains far from its $N \to \infty$ asymptotic behavior (at least per the predictions of \cite{EBKZ-2024-QuenchesSKReluctant}).
Thus, it seems considerably larger experiments would be required to understand whether the final energy reached by reluctant dynamics is indeed non-universal, and we leave such further numerical studies to future work.

\section*{Acknowledgments}
\addcontentsline{toc}{section}{Acknowledgments}

Thanks to Reza Gheissari, Florent Krzakala, and Ilias Zadik for bringing reluctant dynamics to our attention, to Cristopher Moore for helpful discussions, and to the authors of \cite{dandi2025sequential} for pointing out the related results in their paper as well as the predictions of \cite{EBKZ-2024-QuenchesSKReluctant} concerning reluctant dynamics.

\addcontentsline{toc}{section}{References}
\bibliographystyle{alpha}
\bibliography{main}

\end{document}